\documentclass[journal,transmag]{IEEEtran}
\usepackage{float}
\usepackage{wrapfig}
\usepackage{soul}
\usepackage{footnote}
\usepackage{verbatim}
\usepackage{stfloats}
\usepackage{multirow}
\usepackage[dvips]{color}
\usepackage{pifont}
\usepackage{graphicx}

\usepackage[noadjust]{cite}
\renewcommand{\figurename}{Fig.}
\usepackage{subcaption}

\usepackage{amssymb,amstext,amsmath}
\usepackage{float}
\usepackage{booktabs}
\usepackage{array,ragged2e}
\usepackage[table]{xcolor}

\usepackage{multirow}
\usepackage{algorithm}
\usepackage{algpseudocode}
\usepackage{pifont}
\usepackage{enumerate}

\begin{document}

\title{Change Point Models for Real-time Cyber Attack Detection in Connected Vehicle Environment}

\author{\IEEEauthorblockN{Gurcan Comert\IEEEauthorrefmark{1},
Mizanur Rahman\IEEEauthorrefmark{2},
Mhafuzul Islam\IEEEauthorrefmark{2}, and 
Mashrur Chowdhury\IEEEauthorrefmark{2},
\IEEEauthorblockA{\IEEEauthorrefmark{1}Computer Science, Physics, and Engineering, Benedict College, Columbia, SC 29204}
\IEEEauthorblockA{\IEEEauthorrefmark{2}Glenn Department of Civil Engineering, Clemson University, Clemson, SC 29634}
\thanks{Manuscript received January 31, 2020; Corresponding author: G. Comert (email: gurcan.comert@benedict.edu). G. Comert is with the Department of Computer Science, Physics, and Engineering, Benedict College, Columbia, SC 29204, USA. M. Rahman, M. Islam, and M. Chowdhury are with the Glenn Department of Civil Engineering, Clemson University, Clemson, SC 29634, USA. Email: \{mdr, mdmhafi, mac\}@clemson.edu.}
}}

 %
\markboth{IEEE TRANSACTIONS ON INTELLIGENT TRANSPORTATION SYSTEMS}%
{Comert \MakeLowercase{\textit{et al.}}: IEEE Transactions on TITS Journal}

\IEEEtitleabstractindextext{%

\begin{abstract}
Connected vehicle (CV) systems are cognizant of potential cyber attacks because of increasing connectivity between its different components such as vehicles, roadside infrastructure, and traffic management centers. However, it is a challenge to detect security threats in real-time and develop appropriate or effective countermeasures for a CV system because of the dynamic behavior of such attacks, high computational power requirement, and a historical data requirement for training detection models. To address these challenges, statistical models, especially change point models, have potentials for real-time anomaly detections. Thus, the objective of this study is to investigate the efficacy of two change point models, Expectation Maximization (EM) and two forms of Cumulative Summation (CUSUM) algorithms (i.e., typical and adaptive), for real-time V2I cyber attack detection in a CV Environment. To prove the efficacy of these models, we evaluated these two models for three different type of cyber attack, denial of service (DOS), impersonation, and false information, using basic safety messages (BSMs) generated from CVs through simulation. Results from numerical analysis revealed that EM, CUSUM, and adaptive CUSUM could detect these cyber attacks, DOS, impersonation, and false information, with an accuracy of (99\%, 100\%, 100\%), (98\%, 100\%, 100\%), and (100\%, 98\%, 100\%) respectively.
\end{abstract}

}

\maketitle

\IEEEdisplaynontitleabstractindextext

\IEEEpeerreviewmaketitle
\section{Introduction}
\label{intro}
\IEEEPARstart{T}he driving force behind the US economic engine is the surface transportation system, which enables reliable and efficient transportation of passengers and goods~\cite{dot2015beyond}. However, human errors (e.g., poor judgment, fatigue) are the leading causes of more than $94\%$ of US highway fatalities \cite{national2008national}. To reduce these fatalities and associated societal costs by reducing or eliminating the influence of the human errors, the US Department of Transportation (USDOT) has been promoting connected and automated vehicles (CAV)~\cite{USDOT2016,policy2016accelerating}. From recent reports of National Highway Traffic Safety Administration ~\cite{NHTSA2013,NHTSA2016}, several benefits are foreseen with this CAV technologies, such as up to $80\%$ reduction in fatalities from multi-vehicle crashes and preventing the majority of human error related incidents. In such CAV systems, massive amounts of data will be produced and exchanged between different components through different data communication medium, such Dedicated Short Range Communication (DSRC), WiFi, 5G and Long Term Evolution (LTE)~\cite{burt2014big,cvria}. These data can be processed in a cloud, or in an edge computing device at the roadside (i.e., roadside transportation infrastructure) based on different CAV application requirements~\cite{cvria,whaiduzzaman2014survey}. Communication technologies supporting data exchange must also be secured to support CAV operations with specific requirements (e.g., delay, bandwidth and communication range). With the increase of connectivity in transportation networks, this CAV systems is cognizant of potential cyber attacks~\cite{raya2007securing,USGAO}.

As cybersecurity attacks are dynamic, it is a challenge to detect security threats in real-time and develop appropriate or effective countermeasures for connected transportation system \cite{nicolppt}. To increase security and resiliency due to possible attacks or benign system errors by different events, research is needed to investigate detection techniques for different attack types, such as denial of service (DOS), impersonation, false information~\cite{pathre2013novel,mejri2014survey}. Anomaly detection techniques are well-studied in different areas. Specifically, the cybersecurity of firmware updates, cybersecurity on heavy vehicles, vehicle-to-vehicle (V2V) communication interfaces, and trusted vehicle-to-everything (V2X) communications \cite{petit2015potential}.

Different type of anomaly detection models exist in literature, such as rule-based, machine learning (ML) and data mining (DM) (including expert systems)-based, and statistical inference-based models. These can be listed as k-means, random forest, Bayesian networks, Gaussian processes, decision trees, neural networks, support vector machines, and hypothesis testing and point estimation based process control models respectively. Recent survey studies related to anomaly detection are summarized a comprehensive review of machine learning and rule (signature)-based methods, and their applications to intrusion detection systems (IDS)~\cite{buczak2016survey,patcha2007overview}. Rule-based attack detection models, originated from cryptography, are abundant especially for their efficiency and computationally light-weight~\cite{sedjelmaci2014efficient}. However, rule-based models require a detailed understanding of the data generation process and adaptivity or customization based on their respective environment to develop the model. On the other hand, both ML and DM-based attack detection models are adaptable to different attack types both known and unknown patterns \cite{van2019real}. However, major concerns are computational complexity for real-time application, training the model with different cyber attack scenarios, unavailability of cyber attack data in the transportation domain, and determination of update or retraining window. To address these problems, statistical models, specially the change point models, are applicable because of the following advantages: (1) do not require fitting or training; (2) adaptive to different attack data (do not use rules); (3) perform with low data sample sizes; and (4) computationally efficient for real-time applications. Thus, the objective of this study is to investigate the efficacy of two change point models, Expectation Maximization (EM) and Cumulative Sum (CUSUM), for real-time V2I cyber attack detection in a connected vehicle (CV) Environment. To prove the efficacy of these models, we implemented three different type of cyber attacks (i.e., denial of service (DOS), impersonation, and false information)~\cite{van2016survey}, using BSMs generated from CVs through simulation.

The paper is organized as follows. Section~\ref{sctrelwork} presents the previous research and the literature on the anomaly detection models. Section~\ref{sctmethod} describes EM and CUSUM algorithms for V2I cyber-attack detection. Section~\ref{sctne} presents the data generation process and evaluation of EM and CUSUM models through numerical analysis and results. Finally, section~\ref{sctconc} summarizes findings and possible future research directions.
 \vspace{-20pt}
\section{Related Work}
\label{sctrelwork}
In this section, we describe past research on statistical models for anomaly detection and cyber attacks in a V2I environment.
 \vspace{-10pt}
\subsection{Statistical Models for Cyber Attack Detection}
Statistical and inference based models in cyber attack or in general detection problem provide adaptability and transferability to different settings and attack types with low computational costs~\cite{buczak2016survey,carl2006denial}. In a very basic approach, detection on process controls using quality control models based on change point algorithms such as CUSUM, and exponentially weighted moving average are utilized \cite{cardenas2011attacks} intrusion monitoring. For detail characteristics of attack models using honeypot-captured cyber attacks are modeled with several time series models \cite{zhan2013characterizing}. Reliability models are also studied for vulnerabilities based on good and bad states simply via nodes' deviations \cite{mitchell2013effect}. They consider persistent, random, and insidious attacks of sensor-actuator nodes with simple sensing, actuating, and networking models. Moreover, model-based attacks usually for power grids are investigated by researchers \cite{sridhar2014model}. Attack (intrusion) models for different control systems and proper modeling for moving systems as in vehicular or mobile ad hoc network (VANET/MANET) cases are well reviewed in \cite{mitchell2014survey,li2015art,min2017new,liang2019filter} where reputation management in vehicular networks are suggested. Possible revoking or blacklisting the information contributors are also recognized in similar survey study specifically on cooperative intelligent transportation systems \cite{van2016survey}. 

First proposed by Page~\cite{page1954continuous}, CUSUM is a classical statistical quality and process control method for industrial applications, which is then utilized by many fields such as computer network security particularly for DOS or flooding attacks~\cite{tartakovsky2006detection}, sensor networks, signals and control systems, pipeline break detection to neuronal spike detection~\cite{misiunas2005pipeline,ratnam2003change}. However, it is also heavily employed in intrusion or anomaly detection for cyber attacks~\cite{carl2006denial} for its high true positive rate and low computational cost. In connected vehicles, a recent patented implementation utilizes CUSUM on for vehicle intrusion detection on electronic control units~\cite{cho2016fingerprinting}. On the other hand, EM is used for anomaly detection as its classical meaning of parameter estimation in an analytical attack modeling on power systems~\cite{lee2014cyber}. In this study, both EM and CUSUM are selected as detection algorithms for their online applicability (linear in computational complexity) also observed in~\cite{buczak2016survey}. Both algorithms are adopted to the anomaly detection problem as sequential implementation, compared, and detailed attack data are simulated which are novel in the intrusion detection literature. Both methods contain only low level parameters such as initial underlying distributions parameters (e.g., Normal in this paper) as well as design parameters for CUSUM. Detailed recalibration or update intervals for such parameters are not investigated in this study.
 \vspace{-12pt}
\subsection{V2I Cyber-attacks in a CV Environment}
In the cyber-physical systems (CPS) security literature, recent studies  \cite{petit2015potential,mitchell2014survey,humayed2017cyber,kong2017cooperative,biron2018real, mousavinejad2019distributed}, list possible cyber-attacks and discuss their detection and mitigation techniques. In these studies, abstract cyber-physical models for smart cars are also presented. Possible attacks are criminal, privacy, tracking, profiling, political threats with different structures replay, command (message) injection, false information, impersonation, eavesdropping, and denial of service \cite{biron2018real}. 
For this study, we consider denial of service (DOS), impersonation, and false information attack to evaluate efficacy for EM and CUSUM models. DOS attack in the literature defined as disordering, delaying, or periodically dropping packets to decrease network performance. It consists of flooding (similar to jamming-occupying channel by outsiders) and exhausting the network resources such as bandwidth and computational power. In this study, it is dramatically increasing number of messages so that the RSEs or OBEs are not able to process and overall communication delays increase or become not available. Impersonation (node impersonation or identity theft) attack can be defined as a vehicle can pretend as if it has more than one identity unable to distinguish one or more vehicles by aiming to shape the network, manipulating other vehicle behaviors, incorrect position information etc., hard to detect-network/vehicle ID credentials management. False information attack: aims to manipulate other vehicles with selfish/malicious intent can highly impact and high detection likelihood~\cite{sakiz2017survey}.
Previous research on the vehicular communications discuss possible attacks and their mitigation methods \cite{petit2015potential,van2016survey}. ITS applications require protocols that conflicts with anonymity and privacy requirements and report on quantifying such risks and traffic control under either lost communications based on correct or faulty communication errors. In sum, studies on quick detection of such cases and possible redundant data resources for cost effective control are needed for resiliency on transportation networks.
 \vspace{-10pt} 
\section{Change Point Models}
\label{sctmethod}
In this study, we investigate statistical change point models, Expectation Maximization (EM) and Cumulative Summation (CUSUM), to detect cyber attacks in a V2I environment. We describe these models in the following sub-sections.
\subsection{Expectation Maximization Algorithm}
\label{sctEM}
The Expectation Maximization (EM) algorithm is often used to estimate the parameters of mixture models or models with latent variables~\cite{Dempster77maximumlikelihood,hastie2001elements}. In this research, EM algorithm is utilized for detecting cyber attacks via changes in the process mean. Given $N$ sample points from a mixture of two Normal distributions as in Eq.~(\ref{eqn_mix}), the EM algorithm can be applied to determine the parameters of these two distributions {\boldmath$\theta$}=[$\theta_1=(\mu_1,\sigma_1)$, $\theta_2=(\mu_2,\sigma_2)$, $\pi$] of normal and attack states, respectively. The first step of the EM algorithm specifies initial values for the parameters. In the expectation step, the algorithm computes the responsibilities $\gamma_i$ (i.e., the probability of an observation belonging to $Y_2$, i.e., attack state) for each data point. Using the calculated responsibilities, it then computes the five parameters in the maximization step. The iterations continue until the likelihood function convergences. The convergence of a basic EM algorithm is slow. Simple equations pertaining to the EM are given below. First, the probability density of $Y$ is written as a mixture:  
\begin{equation}
Y=(1-\Delta)Y_1+\Delta Y_2
\label{eqn_mix}  
\end{equation}

where $Y_1\sim N(\mu_1,\sigma_1^2)$, $Y_2\sim N(\mu_2,\sigma_2^2)$, and $\Delta \in {0,1}$ with abnormal data proportion of P$(\Delta=1)=\pi$.
\begin{equation}
g_{Y}(y)=(1-\pi)\phi_{\theta_1}(y)+\pi \phi_{\theta_2}(y)
\label{eqn_EM1}  
\end{equation}

where $\phi_{\theta}(x)$ denotes normal density. For a data set of $N$ points the loglikelihood function can be written as follows:
\begin{eqnarray}
l(\theta,Z)=\sum_{i=1}^{N}{ln[(1-\pi)\phi_{\theta_1}(y_{i})+
\pi\phi_{\theta_2}(y_{i})]}
\label{eqn_EM2}  
\end{eqnarray}

where {\boldmath$\theta$}$=[\theta_1=(\mu_1,\sigma_1)$, $\theta_2 = (\mu_2, \sigma_2)$, and $\pi$] and $Z$ represent the data points. Analytical maximization of Eq. (\ref{eqn_EM2}) is difficult, however, if the observation is known to belong to $Y_2$ (i.e., with latent variable $\Delta_i=1$, otherwise $\Delta_i=0$), the loglikelihood can be written as in Eq.~(\ref{eqn_EM3}) and $\Delta_i=1$s can be estimated by Eq.~(\ref{eqn_EM4}).
\vspace{-5pt}
\begin{eqnarray}
l(\theta;\Delta,Z)=\sum_{i=1}^{N}[(1-\Delta_{i})ln[(1-\pi)\phi_{\theta_1}(y_i)]+\\
\Delta_{i}ln[\pi\phi_{\theta_2}(y_i)]] \nonumber
\label{eqn_EM3}  
\end{eqnarray}
\begin{equation}\gamma_{i}(\theta)=E(\Delta_{i} \mid \theta,Z)=P(\Delta_{i}=1\mid \theta,Z)
\label{eqn_EM4}  
\end{equation}
In sum, given $N$ data points that are assumed to be generated by mixture of two Normal distributions (i.e., normal and abnormal messages per vehicle per second (MVS), messages per vehicle (MVT), and distance), the EM algorithm is applied to determine the distribution parameters and responsibilities. Number of mixtures could be varied for various levels of attacks and impacts. $N$ data points constitute the main input to the algorithm. To see the impact of sample size, prediction performances of EM algorithm with various $N$ values can be checked. The EM algorithm provides the real-time estimation of the process parameters at each time point as well as conditional probabilities of a data point comes from a certain attack or no attack condition which is subsequently used for detection.
\vspace{-5pt}
\subsection{CUSUM Algorithm}
\label{sctCUSUM}
The CUSUM chart or algorithm is commonly used for quality control purposes to detect possible shifts in the mean level of a process. In cyber attack setting, changes within expected level of deduced measures (MVS, MVT, and distance) are targeted. This paper uses tabular version or algorithmic version of the CUSUM rather than control chart. Assume that ${X_i\sim}$ identical independently distributed (i.i.d) with known $(\mu_1,\sigma^2)$ where a new process mean is observed $\mu_2$ after a possible change. Based on statistical hypothesis testing, the log-likelihood ratio is written $s(i)$=$ln(p_{\mu_2}(X_i)/p_{\mu_1}(X_i))$ for $S_t=\sum_{i=1}^{t}{s_i}$ for sample size of $n$, the decision rule is given by
\begin{equation}   
C_t=\begin{cases}0~,~S_t<H~;~H_0~no~change\\ 1,~S_t\geq H;~H_1~change \end{cases}   
\label{eqn_dr}  
\end{equation}  

where, $C_t=S_t-m_t$ and $m_t=[S_i]^{-}_{1\leq i \leq t}$.
\subsubsection{Typical Form}
\label{scttypical}
Basic applications of this algorithm assume that the observations collected before and after the change in the mean level are i.i.d. To detect both positive and negative shifts, the two-sided version of the CUSUM algorithm was used. The algorithm works by accumulating positive and negative deviations from a certain target mean, which is commonly taken to be zero. The positive deviations (values above the target) are indicated with $C_{t}^{+}$, and those that are below the target are indicated with $C_{t}^{-}$. The statistics $C_{t}^{+}$ and $C_{t}^{-}$ are referred to as one-sided upper and lower CUSUMs, respectively~\cite{montgomery2009introduction}. It is shown that the use of the two-sided CUSUM algorithm is equivalent to monitoring the following two sums for a zero-mean process:
\begin{eqnarray}
C_{t}^{+}= [0,C_{t-1}^{+}+X_t-\mu_2-K]^+\\ \nonumber
C_{t}^{-}= [0,-C_{t-1}^{-}-X_t+\mu_2-K]^+\\ \nonumber
\label{eqn_CUSUM}  
\end{eqnarray}
\vspace{-25pt}

where $C_{0}^{+}=0$, $C_{0}^{-}=0$, is the residual or deviation from the mean at time $t$. A shift detection is issued whenever $(C_{t}^{+} \lor C_{t}^{-})>H$. Typical CUSUM is applied for persistent shifts or attacks. With $-C_{t-1}^{-}$ in Eq.~(\ref{eqn_CUSUM}), the algorithm behaves like one-sided and reduces false alarm rate almost $100 \%$. Moreover, in order to employ CUSUM in real-time, once an alarm is issued by the CUSUM algorithm, the mean or intercept of the attack time series observations is estimated and updated with Eq.~(\ref{eqn_mu}) and $C_{t}^{+}, C_{t}^{-}$ values set to zero after every detection.
\begin{equation}   
\mu_2=\begin{cases}\mu_1+K+\frac{C_{t}^{+}}{N^{+}}, C_{t}^{+}>H   \\ \mu_1-K-\frac{C_{t}^{-}}{N^{-}}, C_{t}^{-}>H\end{cases}   
\label{eqn_mu}  
\end{equation}

The CUSUM algorithm are designed by choosing the values of $K$ and $H$. The constant $K$ is called the reference value and $H$ is the decision interval or the threshold. The parameter $K$ is a function of the shift in mean level to be detected by the CUSUM algorithm. The value of $H$ is selected to give the largest in-control average run length
(ARL) consistent with an adequately small out-of-control ARL. These two parameters control the ARL, a standard performance measure for online change-detection algorithms. ARL is the average number of data points that have been observed before an out-of-control signal or alarm is generated. There have been many analytical studies on investigating CUSUM's ARL performance. For example, the conventional CUSUM with $K=\delta\sigma/2$ is optimal in detecting a shift of $\delta\sigma$ from target mean. Based on past studies, Montgomery~\cite{montgomery2009introduction} suggests that selecting $K = \delta\sigma/2=\sigma/2$ for $\delta=1$ and $H=5\sigma$ provides a CUSUM algorithm that has good ARL properties against small shifts in the process mean~\cite{montgomery2009introduction}.

The CUSUM algorithm described previously is applied to the change point detection of the time series within basic safety messages. The CUSUM parameters were selected as suggested in the literature: $K = \delta\sigma/2$ and $H = 5\sigma$ and $\delta=1.0$ which represents midpoint between normal and abnormal process means. 
\subsubsection{Adaptive Form}
\label{sctadaptive}
Adaptive version, denoted as aCUSUM, is actually adopted from~\cite{siris2006application} revised to perform for other than zero mean processes, lower false positives, and single weight parameter ($\alpha$). Table~\ref{par_tab} shows only initial mean values are different which could be used as simple as $1^{st}$ value observed in the process. It is applied to $\tilde{X}_t=X_t-\bar{\mu}_{t-1}$. 
\begin{eqnarray}
C_{t}^{+}=[0,C_{t-1}^{+}+\frac{\alpha D_t}{\sigma^2}[X_t-D_t-\alpha D_t/2]]^+\\ \nonumber
C_{t}^{-}=[0,C_{t-1}^{-}-\frac{\alpha D_t}{\sigma^2}[X_t+D_t+\alpha D_t/2]]^+\\ \nonumber
\label{eqn_aCUSUM}  
\end{eqnarray}
\vspace{-25pt}

where, $D_t=(\bar{\mu}_{t}-\mu_{1})$ and $\bar{\mu}_{t}=\alpha \bar{\mu}_{t-1}+(1-\alpha)X_t$.       
This adaptive form of CUSUM algorithm is not very sensitive to $K = \delta\sigma/2$ and $\delta=1.0$. As in the typical algorithm, for less false positive detection $H$ is set to $5\sigma$.
\vspace{-5pt} 
\section{Numerical Experiments}
\label{sctne}
This section presents the data generation to evaluate the methods for different V2I attacks and numerical results.
\vspace{-5pt}
\subsection{Data Generation for V2I Cyber-attacks}
\label{sctsimulation}
In this subsection, data generation process for different type of V2I Cyber-attacks using microscopic traffic simulator is presented. In order to generate the realistic roadway traffic behavior, a microscopic roadway traffic simulation software, Simulation of Urban Mobility (SUMO) is utilized~\cite{krajzewicz2007simulation}. To mimic real-world vehicular movement in a connected vehicle environment, a Roadside Unit (RSU) is assumed to be placed at the Jervy Gym location of Perimeter Road in Clemson, SC, USA. Each vehicle on this roadway are DSRC communication-enabled and can broadcast a part of BSMs (e.g., time stamp, car ID, latitude, longitude and speed) every one-tenth of a second to the RSU.  All vehicle movements data (i.e., BSMs) are recorded in trace files. A trace file is text file that contains time stamp, vehicle ID, latitude, longitude and speed of each vehicle moving on the Perimeter Road, Clemson, SC, USA. The simulation is comprised of $200$ vehicles per hour per lane on Perimeter Road, a four-lane arterial roadway (two lanes each direction) with $35$ miles per hour (mph) speed limit. 
\begin{figure*}[h!]
\centering
\begin{subfigure}{.78\textwidth}
 \centering
\includegraphics[width=0.78\linewidth]{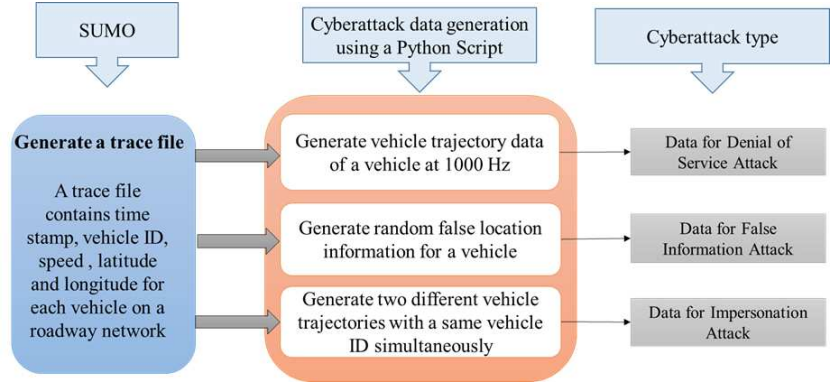}
  \caption{Data generation steps for V2I cyber-attacks.}
  \label{fig_sim}
\end{subfigure}
\begin{subfigure}{.78\textwidth}
\centering
  \includegraphics[width=0.78\linewidth]{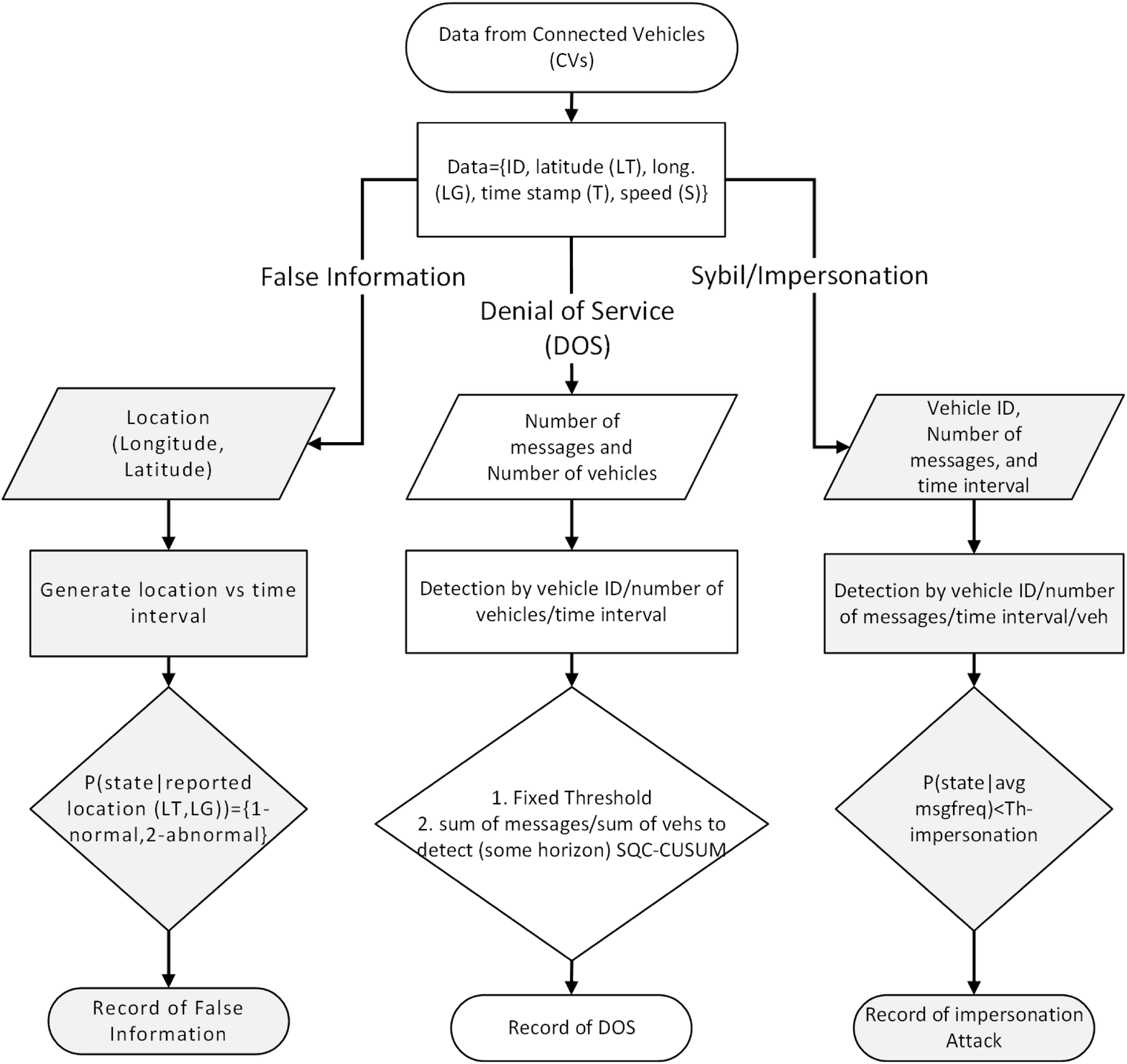}
\caption{Attack detection approach in V2I connected vehicle environment.}
  \label{fig_flow}
\end{subfigure}%
\caption{Data generation steps and attack detection approach}
\label{fig_data}
\end{figure*} 

Using the generated trace file from the SUMO simulation, three different cyber-attack scenarios are generated (see Fig.~\ref{fig_sim}): 
\begin{enumerate}[(i)]
    \item Denial of service (DOS) attack: DSRC has seven communication channels using different frequencies ranging from 5.90 GHz to 5.97 GHz. These seven channels are divided into two categories: Control Channel (CCH) and Service Channel (SCH). Channel number 178 is assigned for CCH and channels 172-182 are assigned as SCH. After the initial authentication and key exchange, the RSU and OBU of a vehicle agree to communicate on a single service channel or frequency. Then, a vehicle can launch the DOS attack by flooding the communication channel in order to cause the service to be unavailable to other vehicles. Typically, an attacker uses its maximum transmission capacity to flood the network. In order to create a breakdown of V2I communication, attackers need to transmit more data than the receiver’s (e.g., RSU) maximum receiving capacity. For generating DOS attack data in our experiment, vehicle with ID $6$ is flooding at 1000Hz while other vehicles are sharing data at 10Hz to mimic the real-world CV environment where each CV is broadcasted BSMs every one-tenth of a second. The total simulation time is $200$ seconds (s) for generating DOS attack data.
    
    \item False information attack: For fake (or false) information attack, false GPS location information (i.e., longitude and latitude) of vehicle ID $2$ are generated simply using random variable generation library from python. We have crafted the attack such way that it creates random location within a given geo-fenced region so that it seems normal geolocation to human. This false information is also broadcasted by the attacker vehicle at 10 Hz or 10 packets/sec. The total simulation time is 200 seconds for false information attack.
    
    \item Impersonation attack:   To emulate the data for impersonation attack, a false ID for vehicle number $3$ is used as vehicle ID $2$. Two different GPS location and speed information for the vehicle ID $2$ are simultaneously generated. In the trace file, the vehicle id of vehicle 3 was replaced by the vehicle id 2 to craft an impersonation attack, where we assume that both of the vehicle 2 and vehicle 3 are in the same region. Thus two different GPS location and speed information are being broadcasted containing the same vehicle id simultaneously. Both of the vehicles is broadcasting the data at 10 packets/sec, and simulation was run for 200 seconds. 
\end{enumerate}

Examples of generated attack data are given in Table~\ref{data_tab}. Evident from the table, multilevel attack monitoring could be designed by vehicle ID and timestamps as micro level tracing (0.1 sec) of such values. However, this approach considerably slows detection capability within time interval of $0.1$ seconds (s) which is critical for safety applications. Therefore, this study tracks aggregate measures such as average message frequency per vehicle per second ($MVS$), average message frequency per vehicle per time interval ($MVT$), distances, and/or track of vehicle speeds within time series framework and detects changes. Detailed vehicle information are not tagged, however, signature is present in the historical data can be traced back for mitigation efforts.
\begin{table}[h!]
\centering
\caption{Examples of attack data generated on RSE}
\scalebox{0.8}{
\label{data_tab}
\begin{tabular}{lrrrrrrrr}
 Type & TS(s) & ID & Lat. & Long. & Speed(m/s) & Pos.(m) & MsgRate \\ 
  \hline
 DOS& 5.10 &   1 & -82.85 & 34.68 & 9.94 & 0.08 & 10.00 \\ 
     & 5.10 &   2 & -82.85 & 34.68 & 8.22 & 0.52 & 10.00  \\ 
     & 5.10 &   3 & -82.85 & 34.68 & 6.21 & 0.74 & 10.00  \\ 
     & 5.10 &   5 & -82.84 & 34.68 & 2.32 & 0.14 & 10.00  \\ 
     & 5.10 &   6 & -82.84 & 34.68 & 0.00 & 0.00 & 10.12  \\ 
     & 5.10 &   6 & -82.84 & 34.68 & 0.00 & 0.00 & 10.23  \\ 
     & 5.10 &   6 & -82.84 & 34.68 & 0.00 & 0.00 & 10.35  \\ 
    \hline
    IMP &  1.30 &   1 & -82.85 & 34.68 & 2.65 & 0.00 & 1.00  \\ 
    &   1.30 &   2 & -82.85 & 34.68 & 0.51 & 0.00 & 1.00  \\ 
    &   1.40 &   1 & -82.85 & 34.68 & 2.87 & 0.00 & 1.00  \\ 
    &   1.40 &   2 & -82.85 & 34.68 & 0.75 & 0.00 & 1.00  \\ 
    &   1.40 &   2 & -82.85 & 34.68 & 1.00 & 0.00 & 2.00 \\ 
    &   1.50 &   1 & -82.85 & 34.68 & 3.19 & 0.00 & 1.00  \\ 
    &   1.50 &   2 & -82.85 & 34.68 & 1.24 & 0.00 & 1.00  \\ 
    \hline
   FAL  & 2.00 &   1 & -82.85 & 34.68 & 4.15 & 0.00 & 1.00  \\
    &   2.00 &   2 & -82.85 & 34.68 & 2.26 & 0.00 & 1.00  \\ 
    &   2.00 &   3 & -82.04 & 34.16 & 0.00 & 72.32 & 1.00  \\ 
    &   2.10 &   1 & -82.85 & 34.68 & 4.31 & 0.00 & 1.00 \\ 
    &   2.10 &   2 & -82.85 & 34.68 & 2.48 & 0.00 & 1.00  \\ 
    &   2.10 &   3 & -82.81 & 34.30 & 0.26 & 71.20 & 1.00  \\ 
    &   2.20 &   1 & -82.85 & 34.68 & 4.57 & 0.00 & 1.00  \\ 
    &   2.20 &   2 & -82.85 & 34.68 & 2.62 & 0.00 & 1.00  \\ 
\noalign{\smallskip}\hline\noalign{\smallskip}
\end{tabular}
}
\end{table}
\vspace{-5pt}
\subsection{Attack Detection Framework}
\label{sctdet}
Fig.~\ref{fig_flow} depicts the approach of attack detection using EM and CUSUM. In order to implement change point detection methods, first step is to identify the processing time window in which information need to track, and how to convert such information in time series behavior to detect shifts due to malicious attacks and/or benign system malfunctions. Such changes result in switching system dynamics and alter critical communications in ITS applications, such as cooperative adaptive cruise control (CACC) and signal control algorithms. In DOS or flooding attacks, vehicles are expected to send more messages than the designed frequency $parameter$ (MVS). Therefore, tracking messages per vehicle and estimating MVS can be used as indicator for cyber-attack detection. For impersonation attack, multiple messages in unit time interval ($0.1$ s) are sent and by monitoring MVT, this type of attack is detected. Lastly, false information attack can be defined as any type of irregularity in the collected messages, such as high or low speed compared to rest of the traffic (inherent) at a roadway segment or an unrealistic gap between any two adjacent vehicles within a certain time frame. CUSUM monitors deviations from process mean and identifies violations. On the other hand, EM calculates conditional probabilities of $P(DOS attack|MVS)>0.001$, where $P(impersonation|MVT)$ and $P(attack state|distance)$ is given. If the likelihoods at any time is $>0.001$, then an attack is detected.  
\vspace{-10pt}
\subsection{Description of EM and CUSUM Parameters}
\label{sctpar}
Parameters for EM and CUSUM are set as provided in Table~\ref{par_tab}. Initialization parameters of EM algorithm are $\theta_1,\theta_2,\pi$, $N=10$ random variates $7$ normal $3$ abnormal, and $10$ iterations per time interval or new observation received. For CUSUM, design parameters as well as initial mean and standard deviations are given in the Table~\ref{par_tab} below. Overall aim here is to give models normal and/or abnormal observations. For instance, in case of DoS attack, $10$ messages per second per vehicle is expected with low or no variations, thus, initial parameters are set to $N(\mu_1=10,\sigma^2=10^{-6})$ for both methods. Moreover, from Table~\ref{attacks_tab}, very small \textit{normal} distance values are calculated from latitude and longitude values (i.e., $\mu_1=0.05$) and false information is calculated to be considerably high so initialized from $N(\mu_2=50,\sigma_{2}^{2}=25)$.       
\begin{table}[h!]
\centering
\caption{Selected model parameters for numerical experiments}
\scalebox{0.75}{
\label{par_tab}
\begin{tabular}{r|l|l|l}
 Type & EM & CUSUM & aCUSUM  \\ 
  \hline
    DOS & $\theta_1=(10,10^{-4}),\theta_2=(15,5),\pi=0.75)$ & $\mu_1=10.00$ &$\mu_1=10.00$\\
    &$Y_{1:7}\sim N(10,10^{-6})$,$Y_{8:10}\sim N(15,10^2)$ & $\sigma=0.001$ & $\sigma=\sqrt{5.10^{-3}\mu_1}$ \\ 
      \hline
  IMP & $\theta_1=(1,10^{-3}),\theta_2=(2,0.5),\pi=0.99)$ & $\mu_1=1.00$ & $\mu_1=1.00$  \\
  
  & $Y_{1:7}\sim N(0.05,10^{-2})$,$Y_{8:10}\sim N(15,10^2)$  & $\sigma=0.01$ & $\sigma=\sqrt{5.10^{-3}\mu_1}$ \\ 
    \hline
    FAL & $\theta_1=(0.05,10^{-2}),\theta_2=(50,5),\pi=0.99)$ & $\mu_1=0.05$ &$\mu_1=1.00$ \\
    
    & $Y_{1:7}\sim N(1,10^{-6})$,$Y_{8:10}\sim N(2,0.5^2)$  & $\sigma=0.1$ &$\sigma=\sqrt{5.10^{-3}\mu_1}$  \\ 
  \hline
 All &$N=10$ and $iteration=10$ & $H=5\sigma$ & $H=5\sigma, K=\delta\sigma/2$ \\   
&  & $K=\delta\sigma/2$ & $\delta=1.00$\\ & & $\delta=1.00$ & $\alpha$=0.025 \\ \hline
\end{tabular}
}
\end{table}
\vspace{-20pt}
\subsection{Analysis and Results}
\label{sctar}
\begin{figure*}[h!]
\centering
\begin{subfigure}{.33\textwidth}
 \centering
\includegraphics[width=1.0\linewidth]{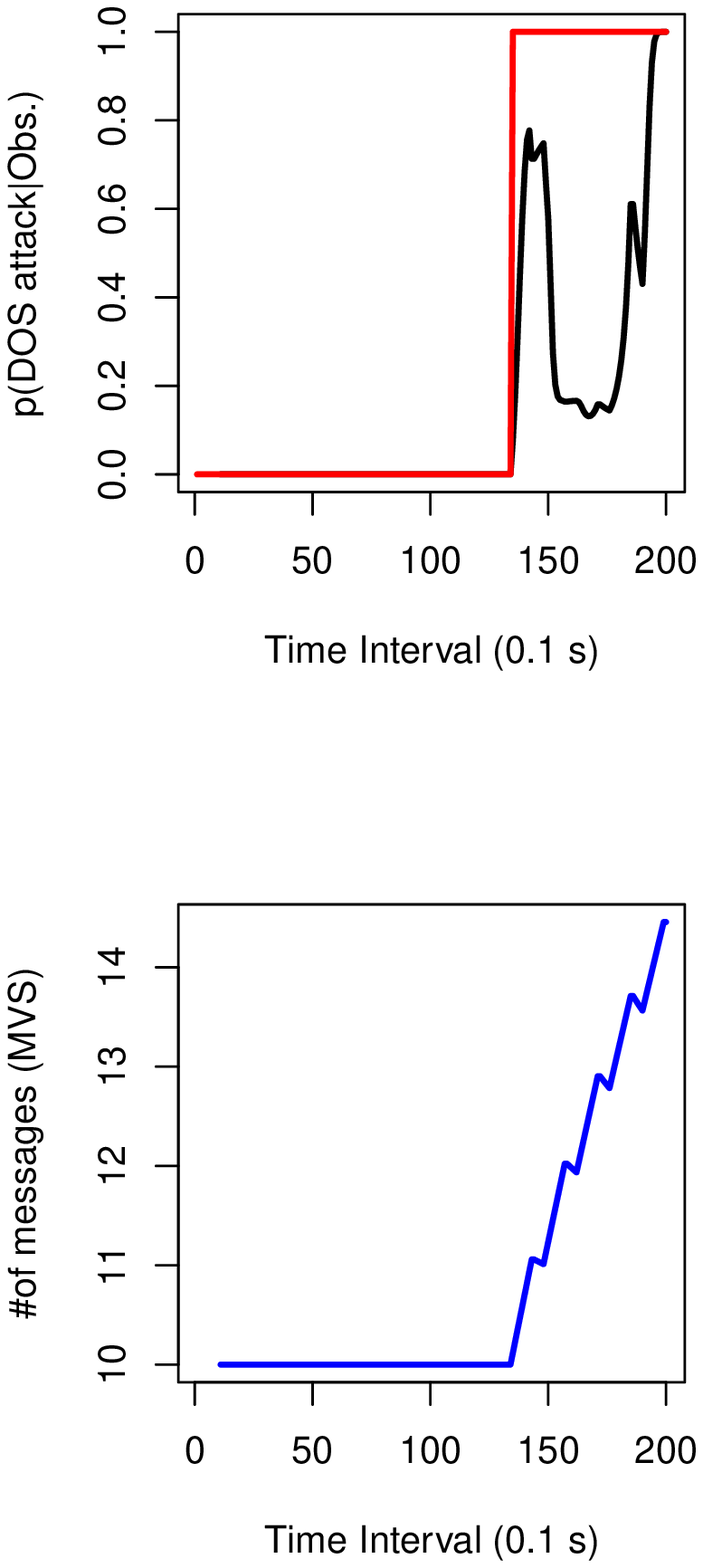}
  \caption{DOS attack}
  \label{fig_dosEM}
\end{subfigure}
\begin{subfigure}{.33\textwidth}
\centering
  \includegraphics[width=1.0\linewidth]{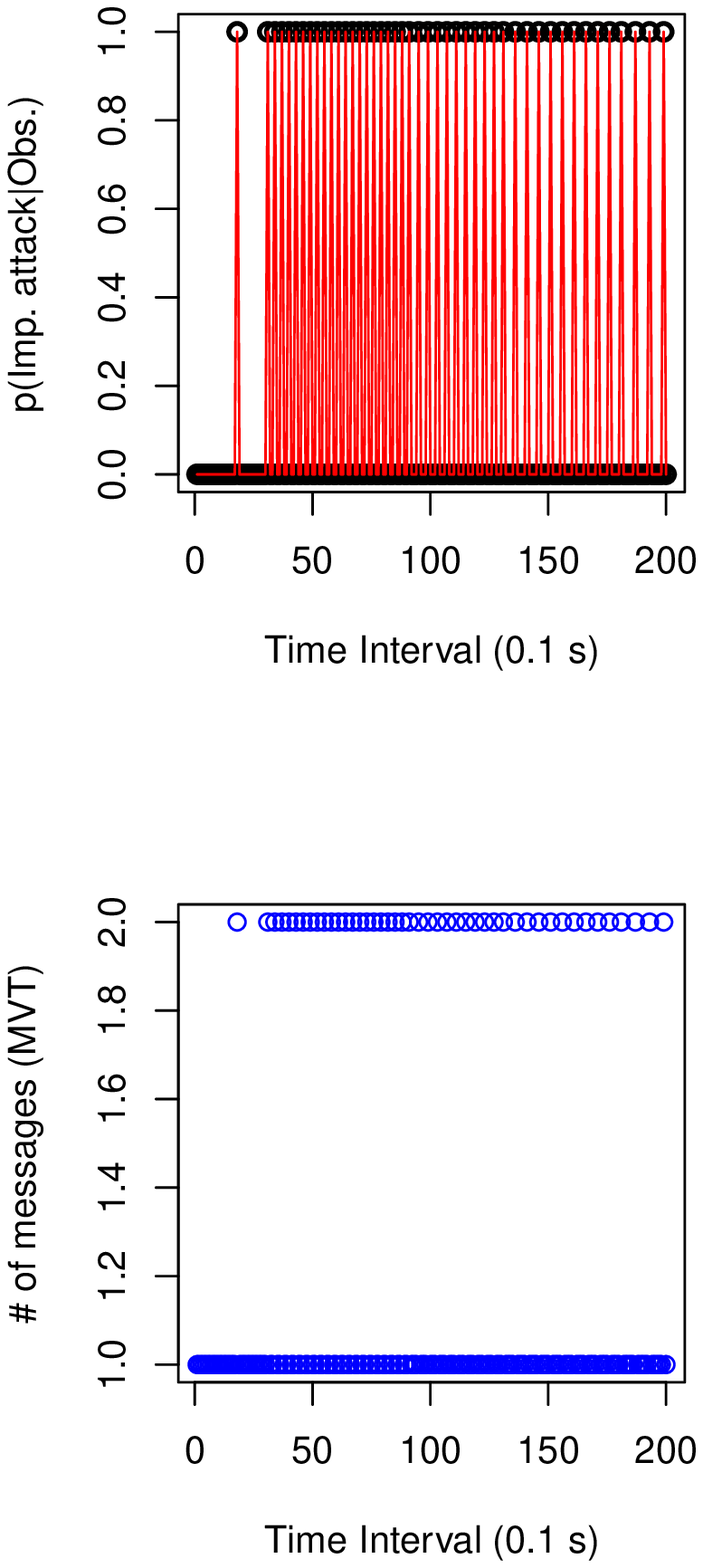}
\caption{Impersonation attack}
  \label{fig_impEM}
\end{subfigure}%
\begin{subfigure}{.33\textwidth}
\centering
  \includegraphics[width=1.0\linewidth]{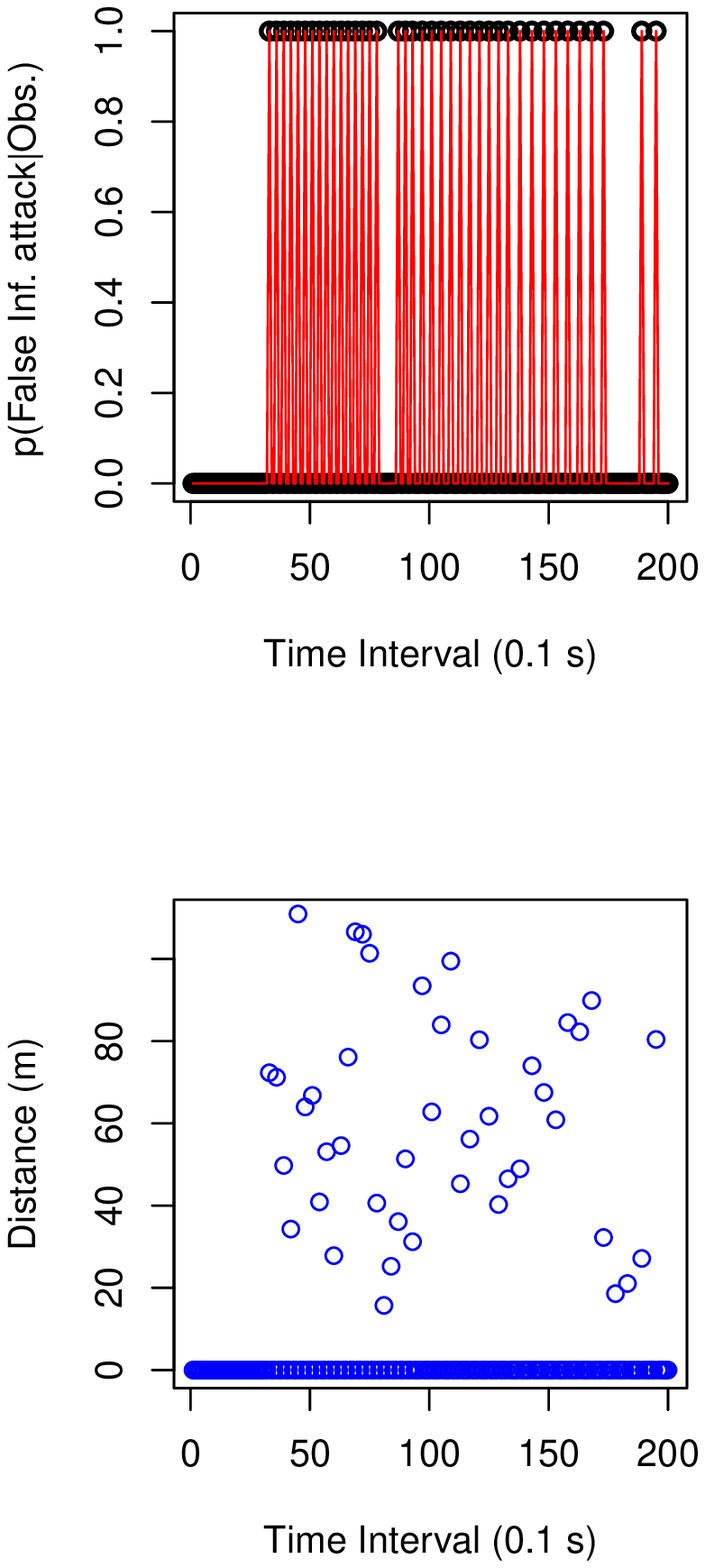}
\caption{False information attack}
  \label{fig_fiEM}
\end{subfigure}%
\caption{Attack detection by EM algorithm}
\label{fig_EM}
\end{figure*}

\begin{figure*}[ht!]
    \centering
            \begin{subfigure}{0.37\textwidth}
            \centering
            \includegraphics[width=1.0\linewidth]{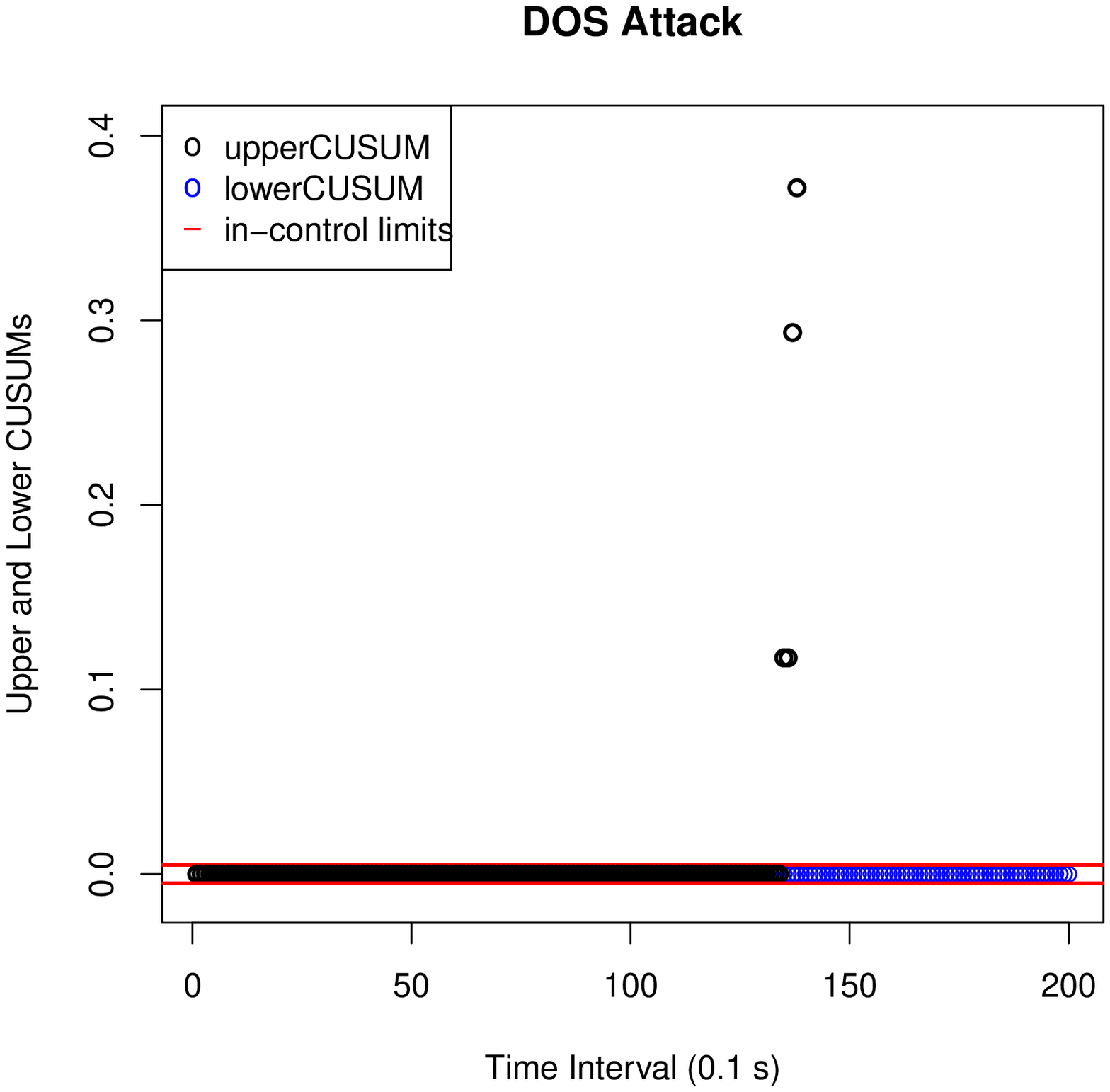}
            \caption{DOS CUSUM}
            \label{fig_cusumdos}
            \end{subfigure}
            \begin{subfigure}{.37\textwidth}
            \centering
            \includegraphics[width=1.0\linewidth]{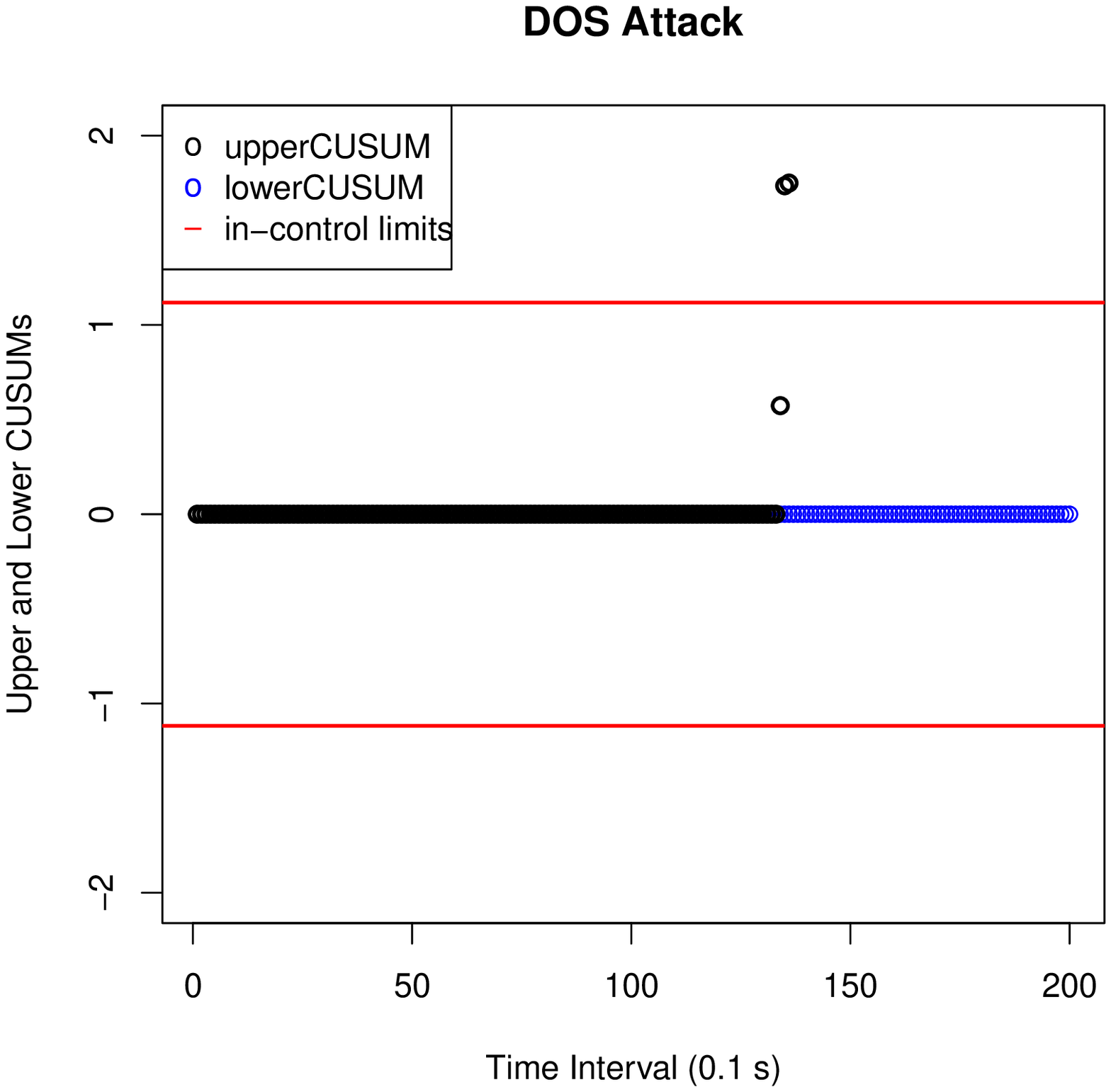}
            \caption{DOS aCUSUM}
            \label{fig_acusumdos}
            \end{subfigure}%
            \\
            \begin{subfigure}{0.37\textwidth}
            \centering
            \includegraphics[width=1.0\linewidth]{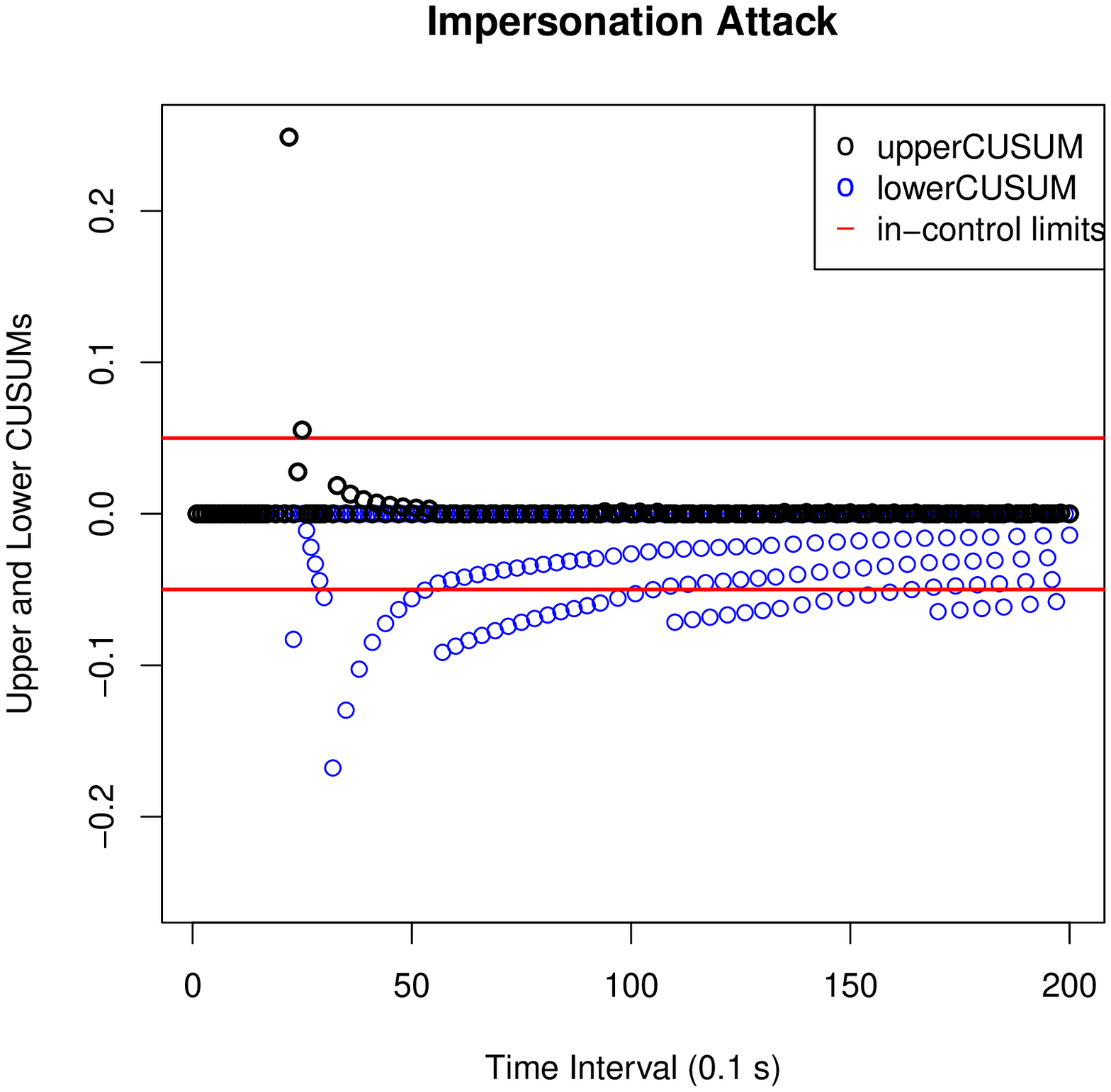}
            \caption{Impersonation CUSUM}
            \label{fig_cusumimp}
            \end{subfigure}
            \begin{subfigure}{0.37\textwidth}
            \centering
            \includegraphics[width=1.0\linewidth]{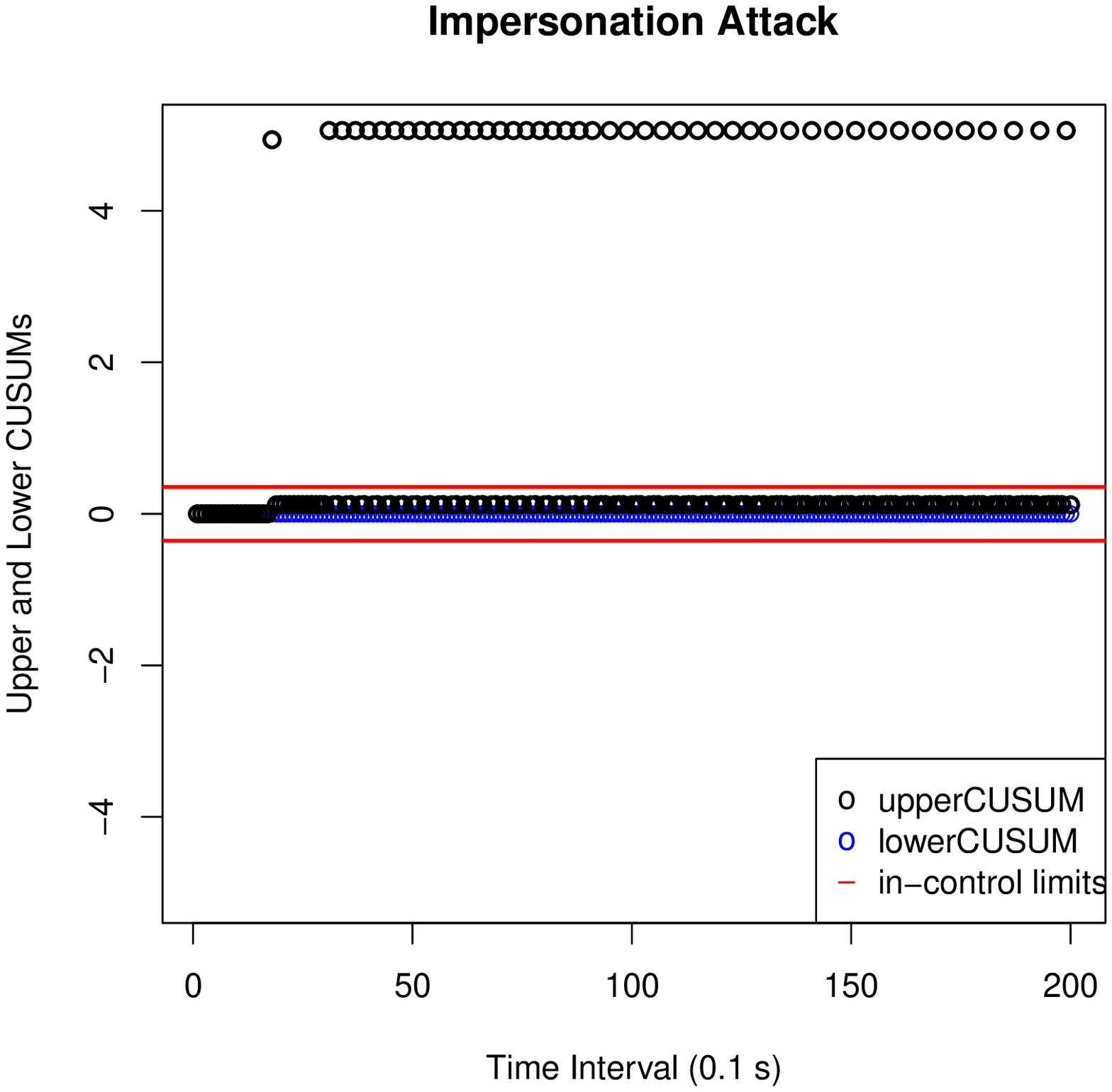}
            \caption{Impersonation aCUSUM}
            \label{fig_acusumimp}
            \end{subfigure}%
            \\   
            \begin{subfigure}{0.37\textwidth}
            \centering
            \includegraphics[width=1.0\linewidth]{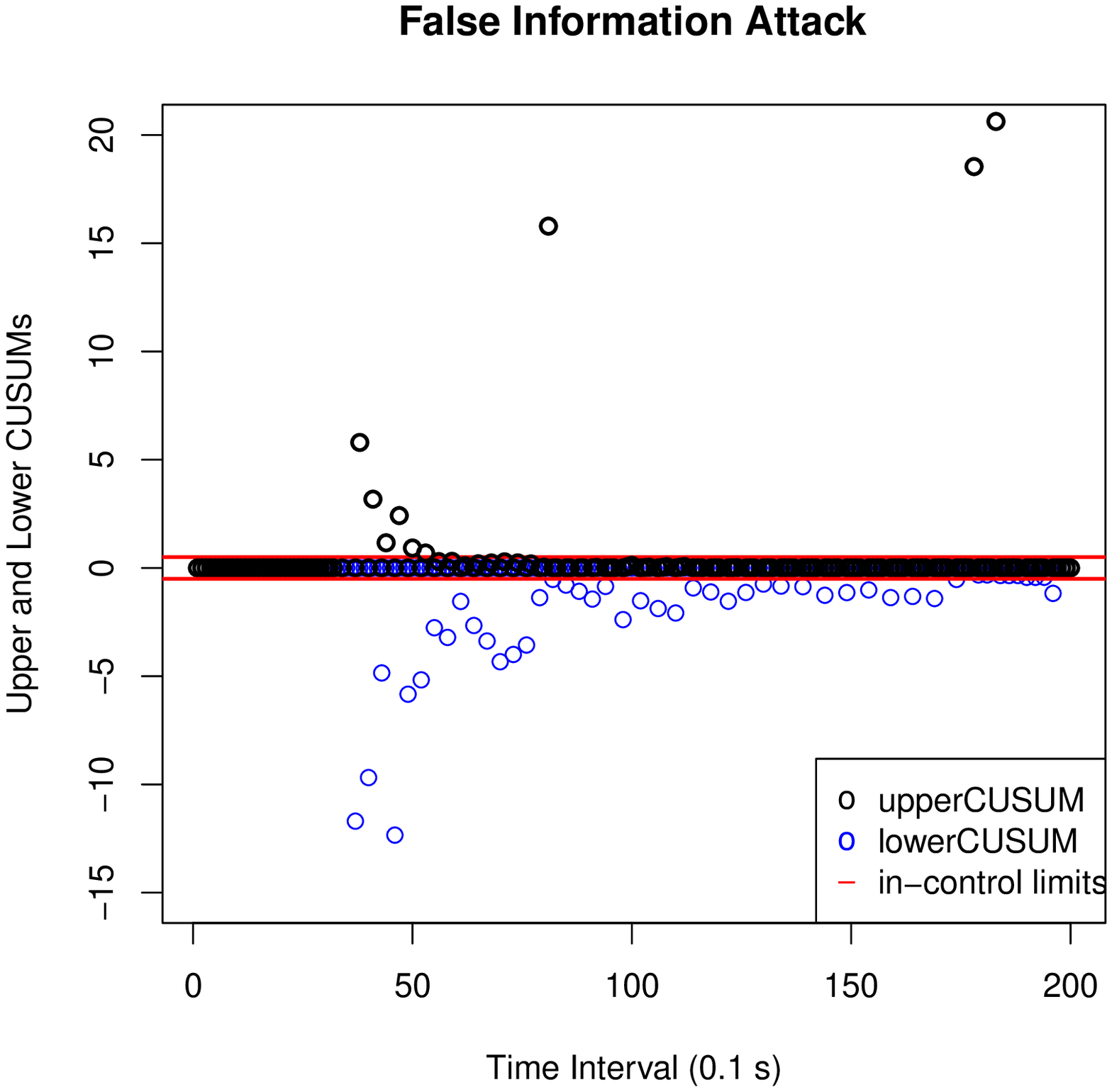}
            \caption{False Information CUSUM}
            \label{fig_cusumfalse}
            \end{subfigure}
            \begin{subfigure}{0.37\textwidth}
            \centering
            \includegraphics[width=1.0\linewidth]{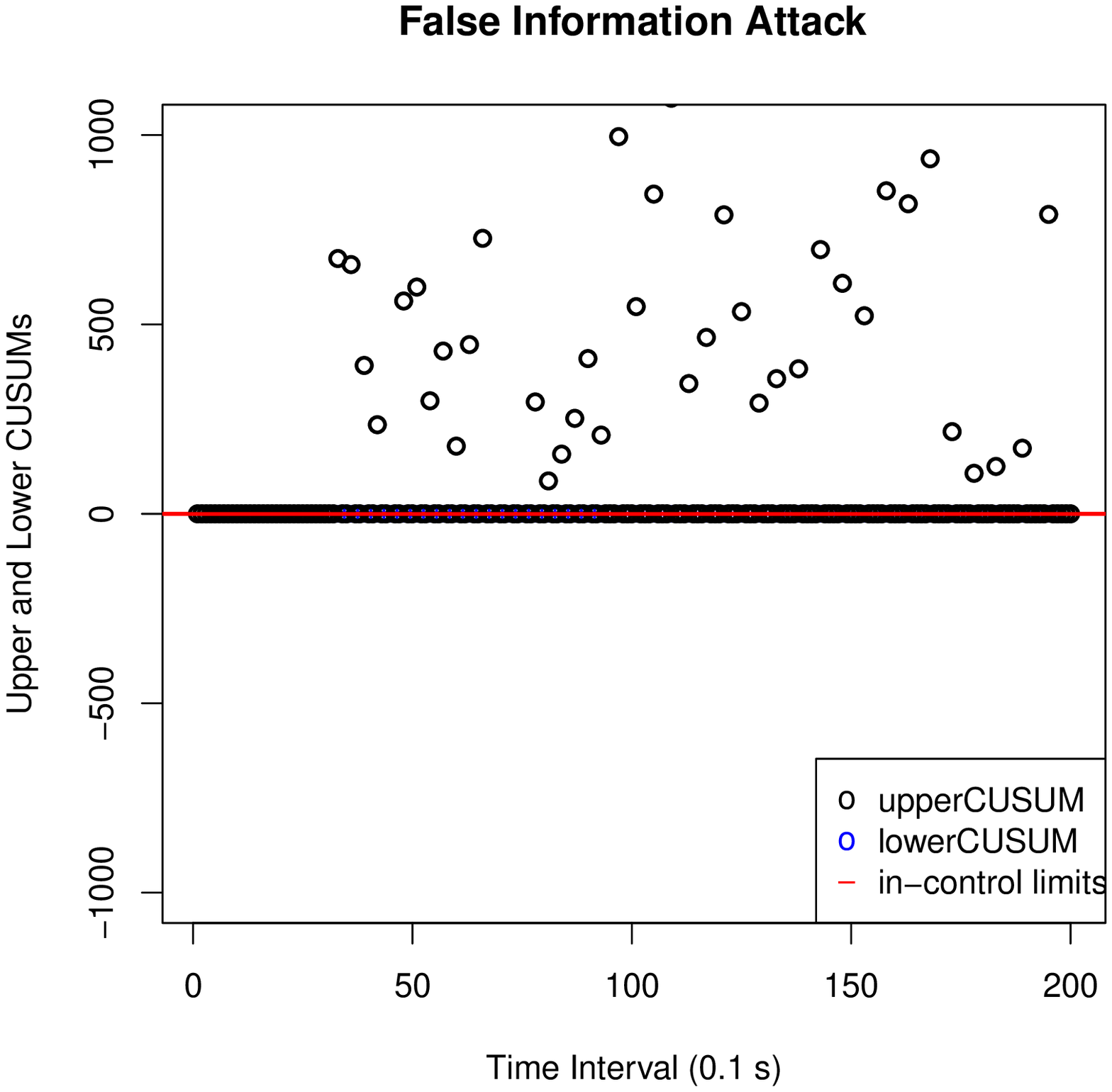}
            \caption{False Information aCUSUM}
            \label{fig_acusumfalse}
            \end{subfigure}%
            \caption{Attack detection by CUSUM algorithm}      
      \label{fig_cusums}
    \end{figure*}

In this section, the effectiveness of attack detection using EM and CUSUM are discussed. Both methods are evaluated using datasets as described in  'Data Generation for V2I Cyber-Attacks' subsection. 
Table~\ref{attacks_tab} provides an example of the generated data from the simulation, attack and detection results. Performances are given as true positive (attack,detected), true negative (no attack, not detected), false positive (no attack, detected) and false negative (attack, not detected) are denoted by TP, TN, FP, and FN respectively. For running the algorithms, we used a PC with 8GB of memory, Pentium I5 Quad-Core CPU. We observed from the table that abnormal behavior is detected accurately by both methods. Since cyber attacks are persistent and CUSUM is based on cumulative differences, shifts are reflected after the detection with rest of observations. Therefore, detection is continuous. This is also evident from Fig.~\ref{fig_cusumdos}-\ref{fig_acusumfalse}. For EM, as a classification algorithm, the detection is based on conditional state probability calculations given the observation and past updated parameters and EM is able to flag normal and abnormal observations (also see Figs.\ref{fig_dosEM}-\ref{fig_fiEM}). Detection alarms are set $P(attack state|observation)>0.001$ and $H=5\sigma$ for EM and CUSUM respectively.  

\begin{figure*}[ht!]
\centering
\includegraphics[width=0.9\linewidth]{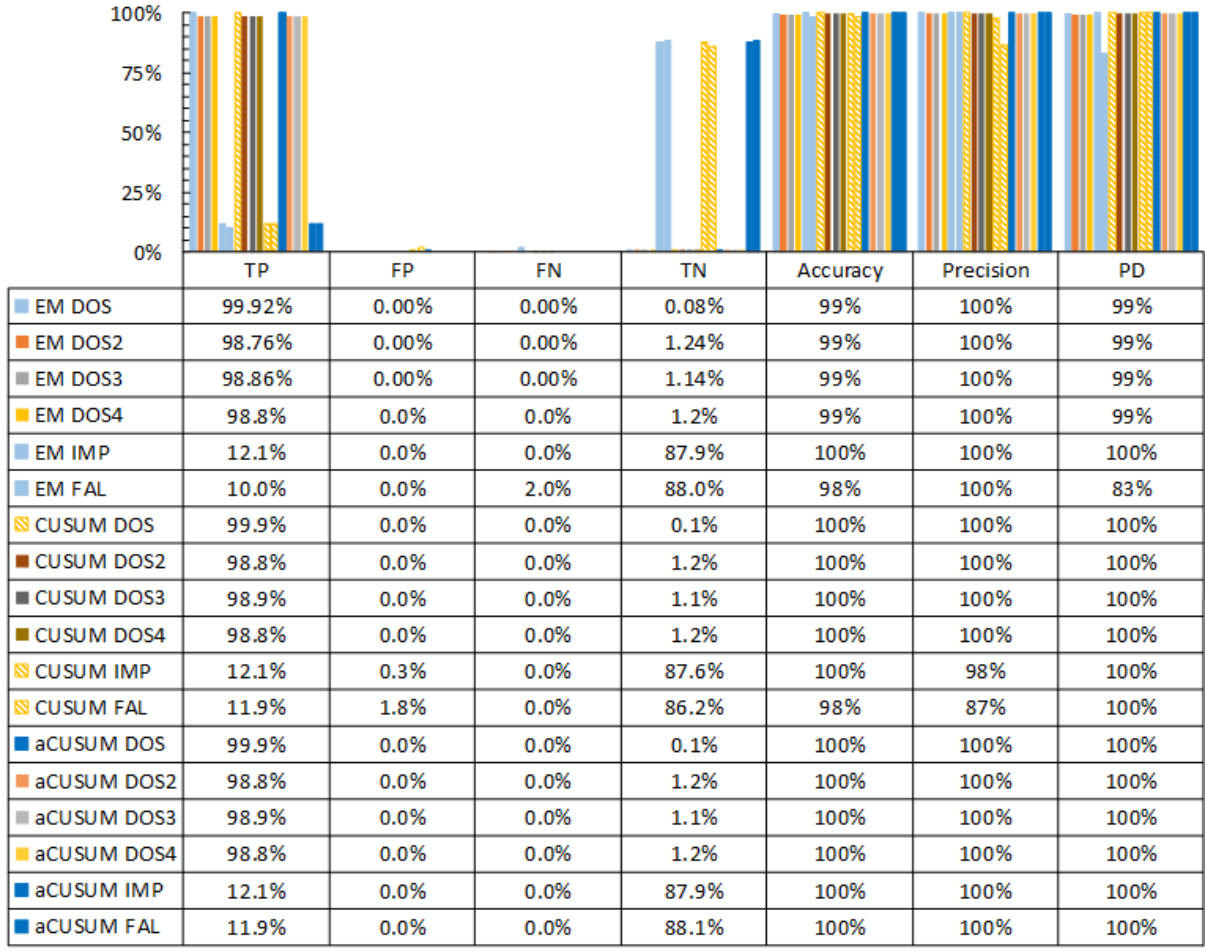}
\caption{Comparison of detection performances between EM and CUSUM algorithms}
\label{fig_acc}
\end{figure*}

\begin{table}[h!]
\centering
\caption{Examples of attack data on RSE and detection by EM and CUSUM}
\label{attacks_tab}
\scalebox{0.65}{
\begin{tabular}{lrrrrrrrlrl}
 Type & Freq. & TS(s) & ID & Spd(m/s) & Pos.(m) & Msgs. & $P(D|Y_t)$ & EM & $(C^+,C^-)$& CUS\\ 
  \hline
DOS & 127 & 5.00 &   1 & 9.73 & 0.08 & 10.00 & 0.00 & TN &(0,0)&TN\\ 
 &128 & 5.00 &   2  & 8.08 & 0.52 & 10.00 & 0.00 & TN &(0,0)&TN\\ 
    &129 & 5.00 &   3 &  5.97 & 0.74 & 10.00 & 0.00 & TN& (0,0)&TN\\ 
     &130 & 5.00 &   5 &  2.14 & 0.40 & 10.00 & 0.00 & TN& (0,0)&TN\\ 
    &131 & 5.10 &   1 & 9.94 & 0.08 & 10.00 & 0.00 & TN &(0,0)&TN\\ 
     &132 & 5.10 &   2 &  8.22 & 0.52 & 10.00 & 0.00 & TN& (0,0)&TN\\ 
     &133 & 5.10 &   3 & 6.21 & 0.74 & 10.00 & 0.00 & TN& (0,0)&TN\\ 
     &134 & 5.10 &   5 &  2.32 & 0.14 & 10.00 & 0.00 & TN& (0,0)&TN\\ 
     &135 & 5.10 &   6 &  0.00 & 0.00 & 10.12 & 0.02 & TP& (0.12,0)&TP\\ 
     &136 & 5.10 &   6 &  0.00 & 0.00 & 10.23 & 0.05 & TP& (0.12,0)&TP\\ 
     &137 & 5.10 &   6 &  0.00 & 0.00 & 10.35 & 0.09 & TP& (0.29,0)&TP\\ 
    \hline
   IMP &  13 & 1.20 &   1  & 2.44 & 0.00 & 1.00 & 0.00 & TN &(0,0) &TN\\
    &  14 & 1.20 &   2 &  0.26 & 0.00 & 1.00 & 0.00 & TN & (0,0)&TN\\ 
    &  15 & 1.30 &   1 &  2.65 & 0.00 & 1.00 & 0.00 & TN & (0,0)&TN\\ 
    &  16 & 1.30 &   2 &  0.51 & 0.00 & 1.00 & 0.00 & TN & (0,0)&TN\\ 
    &  17 & 1.40 &   1 &  2.87 & 0.00 & 1.00 & 0.00 & TN & (0,0)&TN\\ 
    &  18 & 1.40 &   2 &  0.75 & 0.00 & 1.00 & 0.00 & TN & (0,0)&TN\\ 
    &  19 & 1.40 &   2 &  1.00 & 0.00 & 2.00 & 0.67 & TP &0.99,0 &TP\\ 
    &  20 & 1.50 &   1 &  3.19 & 0.00 & 1.00 & 0.00 & TN & 0,0.99&FP\\ 
    &  21 & 1.50 &   2 &  1.24 & 0.00 & 1.00 & 0.00 & TN & 0.99,0&FP\\ 
    \hline
   FAL &  31 & 2.00 &   1  & 4.15 & 0.00 & 1.00 & 0.00 & TN & (0,0) &TN \\
    &  32 & 2.00 &   2 &  2.26 & 0.00 & 1.00 & 0.00 & TN & (0,0)&TN\\ 
    &  33 & 2.00 &   3  & 0.00 & 72.32 & 1.00 & 1.00 & TP & (72.2,0)&TP\\ 
    &  34 & 2.10 &   1  & 4.31 & 0.00 & 1.00 & 0.00 & TN & (0,72.3)&FP\\ 
    &  35 & 2.10 &   2 &  2.48 & 0.00 & 1.00 & 0.00 & TN & (72.2,0)&FP\\ 
    &  36 & 2.10 &   3 &  0.26 & 71.20 & 1.00 & 1.00 & TP & (34.9,0)&TP\\ 
    &  37 & 2.20 &   1 & 4.57 & 0.00 & 1.00 & 0.00 & TN & (0,11.7)&FP\\ 
    &  38 & 2.20 &   2 & 2.62 & 0.00 & 1.00 & 0.00 & TN & (5.8,0)&FP\\ 
    &  39 & 2.20 &   3 &  0.52 & 49.76 & 1.00 & 1.00 & TP & (48.2,0)&TP\\ 
    &  40 & 2.30 &   1 & 4.81 & 0.00 & 1.00 & 0.00 & TN & (0,9.7)&FP\\ 
    &  41 & 2.30 &   2 &  2.83 & 0.00 & 1.00 & 0.00 & TN & (3.2,0)&FP\\ 
    &  42 & 2.30 &   3 &  0.75 & 34.29 & 1.00 & 1.00 & TP & (33.6,0)&TP\\ 
    &  43 & 2.40 &   1 &  4.95 & 0.00 & 1.00 & 0.00 & TN & (0,4.9)&FP\\ 
\noalign{\smallskip}\hline\noalign{\smallskip}
\end{tabular}
}
\footnotesize{true positive (TP), true negative (TN), false positive (FP), false negative (FN)}\\
\end{table}
In Table~\ref{attacks_tab}, position column is calculated in meters (m) from two consecutive latitude and longitude values by using the generic formula: $Pos=1242sin^{-1}(\sqrt{a})$ where $a=0.5-cos((x_2-x_1)p)/2+cos(px_1)cos(px_2)(1-cos((y_2-y_1)p))/2$ and $p=\pi/180$. As discussed above, MST and MSV measures are deduced from time and ID columns for every time interval of $0.1$ s and time series are generated for statistical detection. It should also be noted that for the DOS attack vehicle number $6$ is not sending speed and location correctly.  Attack detection using the change of speed and distance would be trivial. Attacker would also replicate reasonable values. So, detection is carried out using message frequency in MSV. From the table, EM's $P(attack|observation)$ is denoted as $P(D|Y_t)>0.001$ resulting as detection, otherwise no detection. Similarly, for CUSUM $(C^+,C^-)$ values are given. Based on these values, when $(C^+ \lor C^-)>5\sigma$ a detection is observed, otherwise ND is issued. Persistent attacks are easily detected by CUSUM and EM. CUSUM continues to detect normal observations as attacks as an out-of-control process and generates false positive errors. This can be fixed in CUSUM with a slight revision in $C^-$ values mimicking one-sided control. However, in this study, the performance of a typical CUSUM has been investigated without any modifications. EM's performance on false positive errors is promising. Detailed detection performance metrics are presented in Fig.~\ref{fig_acc}.

Figs.~\ref{fig_dosEM}-\ref{fig_fiEM} depict performance of EM algorithm for detecting different attacks. Fig.~\ref{fig_dosEM} shows the likelihood of an attack given $135^{th}$ observation that is also given in \textit{msg} column in Table~\ref{attacks_tab} as $10.12>10.00$, i.e., $P(attack|Y_{135}=10.12)=0.02>0.0001$ (see \textit{EM} column) a very low practical threshold. $P(attack|Y_t)$ increases as frequency values gets larger. For other type of attacks, the changes in observations are not gradual rather sudden which leads to $P(attack|Y_{19}=2)=0.67$ and $P(attack|Y_{33}=72.32)=1.00$ in impersonation and false information attacks, respectively. However, this statistical inference via EM comes with a computational cost. Especially for DOS attack where change is gradual and more messages sent per vehicle, therefore, more data points to be processed per time step ends up with higher computational time. In Table~\ref{comp_tab}, \textit{Attack} column for EM and CUSUM shows computational times of $50$, $200$, and $4761$  to process all the data points. Thus, the feasibility of using EM for DOS attack detection using messages per vehicle per time $1$ s would be second-by-second monitoring. Similarly, false information attack would also require about a second ($0.53$ s). Only, impersonation attack seems feasible to detect within $0.1$ s. These results are consistent with the approximate computational complexity of EM being $O(nkj)$ where $n$ is sample size or time step and $k=2$ is the number of mixtures, and j=$10$ denotes the number of iterations. Similarly, it is linear for CUSUM $O(nm)$ with $m$ being number of elementary operations within each $n$ time interval.  
\begin{table}[h!]
\centering
\caption{Computational times in seconds experienced for EM and CUSUMs}
\label{comp_tab}
\scalebox{0.8}{
\begin{tabular}{ll|lll|lll|lll}
  \multicolumn{2}{c}{Attack Type}&\multicolumn{3}{c}{EM}&\multicolumn{3}{c}{CUSUM}&\multicolumn{3}{c}{aCUSUM} \\ 
   \hline
  & n= & 50 & 200 & 4761 & 50 & 200 & 4761 & 50 & 200 &4761 \\ 
   \hline
    DOS &  &2.19 & 2.24 & 44.58 & 0.43 & 0.46 & 0.65 & 0.77 & 0.77 &1.02  \\ 
    IMP &  &0.02 & 0.02 & 0.02 & 0.01 & 0.01 & 0.02 & 0.01& 0 & 0.01\\ 
    FAL &  &0.53 & 6.15 & 44.91 & 0.03 & 0.06 & 0.19 & 0.14 & 0.14 &0.28\\ 
   \hline
\end{tabular}
}
\end{table}
    
Fig.~\ref{fig_cusums} presents detection results of the CUSUM and aCUSUM algorithms for first 200 data points with $0.1$ s intervals. Shorter intervals are shown in order to provide legibility. In Figs.~\ref{fig_cusumdos}-\ref{fig_acusumdos}, a shift occurs at $135^{th}$ observation for DOS attack. CUSUMs advantage over EM is that it can be implemented for short time intervals due to less computational times. The duration for detecting DOS attack using EM is higher than $0.1$ s interval. However, impersonation and false information attacks can be detected within $0.1$ s (see Figs.~\ref{fig_cusumimp}-\ref{fig_acusumfalse}). Given sufficient time window, EM algorithm would be able to adapt to detect different attack types with new set of normal data set is fed. It has less parameters to be tuned compared to CUSUMs and prone less to false positive alarms. In their simple forms, they are vulnerable to high false positive when adaptive thresholds are used. CUSUMs are very sensitive to real-time estimation or update of $\mu_1,\mu_2,\sigma$ values. In another appropriate midterm application, an hybrid method can be developed to estimate these parameters with EM and input to CUSUMs. Because of space limitations, these experiments are left for another study. 

In Fig.~\ref{fig_acc}, we compared detection performances of the models. Metrics adopted from ~\cite{buczak2016survey} are given as true positive (TP), true negative (TN), false positive (FP) and false negative (FN) are inserted in $accuracy$=$(TP+TN)/(TP+TN+FP+FN)$, $precision$=$TP/(TP+FP)$, $sensitivity$ or $detection$=$TP/(TP+FN)$~\cite{buczak2016survey}. EM only contains about $2\%$ FN for false information and $1\%$ FN for DOS attacks where CUSUM gives $11.8\%$ FP for false information and $2.2\%$ for impersonation attack. For false information attack, EM gives only $83\%$ sensitivity measure and CUSUM is low $87\%$ in precision. After carefully tuning, aCUSUM outperforms both EM and typical CUSUM with no FP and FN for all attack types.  
\vspace{-20pt}
\section{Conclusions}
\label{sctconc}
In this study, we investigated the efficacy of two main statistical change point models, EM and CUSUM, for real-time V2I cyber attack detection in a CV Environment. To prove the efficacy of these models, we evaluated these two models for three different type of cyber attacks, denial of service (DOS), impersonation, and false information, using BSMs generated from CVs. A comprehensive attack modeling is developed for all type of cyber attacks. To generate the data for different cyber attacks, a microscopic traffic simulation software, SUMO, was used for simulating realistic traffic behavior. Instead of tracking data values such as message frequency, speed, and distance individually for each time interval and vehicle ID, aggregate measures are deduced from BSMs to be used in effective real-time detection. Based on the numerical analysis, we found that: 
\begin{enumerate}
\item Given proper initialization, i.e., mean and variance measures of normal and abnormal cases, and enough computational power, both algorithms can detect all three attack types accurately.
\item When attack detection time window is critical such as safety applications, detection time window for EM is greater than  $<0.1$ sec, whereas, the time window for CUSUM is below $0.1$ sec computational times. 
\item When multiple states could be observed for an attack or to classify different impacts, as well as any changes in the normal RSU communication frequencies, EM algorithm would be able to provide conditional probabilities for multiple states.
\end{enumerate}
Results from numerical analysis also revealed that both EM, CUSUM, and aCUSUM could detect these cyber attacks with an accuracy of at least 98\%, 98\%, and 100 \% respectively. Models can be applied for real-time cyber attack detection with a one-second interval. Possible improvements to this research and future directions can be followings:
(1) further research is needed to investigate factors affecting the optimal selection of such parameters with multiple data sets; (2) hybrid methods can be formulated for detection both fast and less sensitive to initialization, and (3) as data generation processes expected to be correlated, algorithms within state-space time series models can be utilized.
\vspace{-15pt}
\section*{Acknowledgments}
This study is partially supported by the Center for Connected Multimodal Mobility ($C^{2}M^{2}$) (USDOT Tier 1 University Transportation Center) headquartered at Clemson University, Clemson, South Carolina. Any opinions, findings, and conclusions or recommendations expressed in this material are those of the authors and do not necessarily reflect the views of the Center for Connected Multimodal Mobility ($C^{2}M^{2}$) and the official policy or position of the USDOT/OST-R, or any State or other entity, and the U.S. Government assumes no liability for the contents or use thereof. It is also partially supported by U.S. Department of Homeland Security Summer Research Team Program Follow-On grant and NSF Grant No. 1719501.
 \vspace{-20pt}
\bibliographystyle{IEEEtran}
\bibliography{detection_trb}
\vspace{-40pt}
\begin{IEEEbiography}{Gurcan Comert} received the B.Sc. and M.Sc.
degree in Industrial Engineering from Fatih University, Istanbul, Turkey and the Ph.D. degree in Civil Engineering from University of South Carolina, Columbia, SC, in 2003, 2005, and 2008 respectively.
He is currently with Computer Science, Physics, and Engineering Department, Benedict College, Columbia, SC. His research interests include applications of statistical models to transportation problems such as traffic parameter prediction, and stochastic modeling.
\end{IEEEbiography}
\vspace{-40pt}
\begin{IEEEbiography}{Mizanur Rahman}
received his Ph.D. and M.Sc. degree in civil engineering with transportation systems major in 2018 and 2013, respectively, from Clemson University. Since 2018, he has been a research associate of the Center for Connected Multimodal Mobility ($C^{2}M^{2}$), a U.S. Department of Transportation Tier 1 University Transportation Center (cecas.clenson.edu/c2m2) at Clemson University. He was closely involved in the development of Clemson University Connected and Autonomous Vehicle Testbed (CU-CAVT). His research focuses on transportation cyber-physical systems for connected and autonomous vehicles and for smart cities.
\end{IEEEbiography}
\vspace{-50pt}
\begin{IEEEbiography}{Mhafuzul Islam} received the BS degree in Computer Science and Engineering from the Bangladesh University of Engineering and Technology in 2014 and MS degree in Civil Engineering from Clemson University at 2018. He is currently a Ph.D. student in the Glenn Department of Civil Engineering at Clemson University. His research interests include Transportation Cyber-Physical Systems with an emphasis on Data-driven Connected Autonomous Vehicle. He is a student member of IEEE.
\end{IEEEbiography}
\vspace{-50pt}
\begin{IEEEbiography}{Mashrur Chowdhury}{Mashrur Chowdhury} (SM'12) received the Ph.D. degree in civil engineering from the University of Virginia, USA in 1995. Prior to entering academia in August 2000, he was a Senior ITS Systems Engineer with Iteris Inc. and a Senior Engineer with Bellomo McGee Inc., where he served as a Consultant to many state and local agencies, and the U.S. Department of Transportation on ITS related projects. He is the Eugene Douglas Mays Professor of Transportation with the Glenn Department of Civil Engineering, Clemson University, SC, USA. He is also a Professor of Automotive Engineering and a Professor of Computer Science at Clemson University. He is the Director of the USDOT Center for Connected Multimodal Mobility ($C^{2}M^{2}$) (a TIER 1 USDOT University Transportation Center). He is Co-Director of the Complex Systems, Data Analytics and Visualization Institute (CSAVI) at Clemson University. He is also the Director of the Transportation Cyber-Physical Systems Laboratory at Clemson University. He serves as an Associate Editor for the IEEE TRANSACTIONS ON INTELLIGENT TRANSPORTATION SYSTEMS and Journal of Intelligent Transportation Systems. He is a Fellow of the American Society of Civil Engineers and a Senior Member of IEEE.
\end{IEEEbiography}

\end{document}